\documentclass[manuscript]{acmart}
\AtBeginDocument{%
  }

\setcopyright{acmlicensed}
\copyrightyear{2018}
\acmYear{2018}
\acmDOI{XXXXXXX.XXXXXXX}
\acmConference[Conference acronym 'XX]{Make sure to enter the correct
  conference title from your rights confirmation email}{June 03--05,
  2018}{Woodstock, NY}
\acmISBN{978-1-4503-XXXX-X/2018/06}

\begin{document}

%%
%% The "title" command has an optional parameter,
%% allowing the author to define a "short title" to be used in page headers.
\title{Messages in a Digital Bottle: A Youth-Coauthored Perspective on LLM Chatbots and Adolescent Loneliness}

%%
%% The "author" command and its associated commands are used to define
%% the authors and their affiliations.
%% Of note is the shared affiliation of the first two authors, and the
%% "authornote" and "authornotemark" commands
%% used to denote shared contribution to the research.
\author{Jinyao Liu}
\affiliation{%
  \institution{Wycombe Abbey}
  \city{High Wycombe}
  \country{UK}}
\email{dora.jinyaoliu@gmail.com}

\author{Di Fu}
\authornote{Corresponding author.}
\affiliation{%
  \institution{School of Psychology, Surrey Institute for People-Centred Artificial Intelligence, University of Surrey}
  \city{Guildford}
  \country{UK}}
\email{d.fu@surrey.ac.uk}

%%
%% By default, the full list of authors will be used in the page
%% headers. Often, this list is too long, and will overlap
%% other information printed in the page headers. This command allows
%% the author to define a more concise list
%% of authors' names for this purpose.
\renewcommand{\shortauthors}{Liu et al.}

%%
%% The abstract is a short summary of the work to be presented in the
%% article.
\begin{abstract}
Adolescent loneliness is a growing concern in digitally mediated social environments. This work-in-progress presents a youth-authored critical synthesis on chatbots powered by Large Language Model (LLM) and adolescent loneliness. The first author is a 16-year-old Chinese student who recently migrated to the UK. She wrote the first draft of this paper from her lived experience, supervised by the second author. Rather than treating the youth perspective as one data point among many, we foreground it as the primary interpretive lens, grounded in interdisciplinary literature from social computing, developmental psychology, and Human-Computer Interaction (HCI). We examine how chatbots shape experiences of loneliness differently across adolescent subgroups, including those with
anxiety or depression, neurodivergent youth, and immigrant adolescents, and identify both
conditions under which they may temporarily reduce isolation and breakdowns that risk
deepening it. We derive three population-sensitive design implications. The next phase of
this work will expand the youth authorship model to a panel of adolescents across these
subgroups, empirically validating the framework presented here.
\end{abstract}

%%
%% The code below is generated by the tool at http://dl.acm.org/ccs.cfm.
%% Please copy and paste the code instead of the example below.
%%
%%\begin{CCSXML}
%%<ccs2012>
%% <concept>
%%  <concept_id>00000000.0000000.0000000</concept_id>
%%  <concept_desc>Do Not Use This Code, Generate the Correct Terms for Your Paper</concept_desc>
 %% <concept_significance>500</concept_significance>
 %%</concept>
%% <concept>
 %% <concept_id>00000000.00000000.00000000</concept_id>
 %% <concept_desc>Do Not Use This Code, Generate the Correct Terms for Your Paper</concept_desc>
 %% <concept_significance>300</concept_significance>
 %%</concept>
 %%<concept>
 %% <concept_id>00000000.00000000.00000000</concept_id>
 %% <concept_desc>Do Not Use This Code, Generate the Correct Terms for Your Paper</concept_desc>
 %% <concept_significance>100</concept_significance>
 %%</concept>
%% <concept>
 %% <concept_id>00000000.00000000.00000000</concept_id>
 %% <concept_desc>Do Not Use This Code, Generate the Correct Terms for Your Paper</concept_desc>
 %% <concept_significance>100</concept_significance>
%% </concept>
%%</ccs2012>
%%\end{CCSXML}

%%\ccsdesc[500]{Do Not Use This Code~Generate the Correct Terms for Your Paper}
%%\ccsdesc[300]{Do Not Use This Code~Generate the Correct Terms for Your Paper}
%%\ccsdesc{Do Not Use This Code~Generate the Correct Terms for Your Paper}
%%\ccsdesc[100]{Do Not Use This Code~Generate the Correct Terms for Your Paper}

%%
%% Keywords. The author(s) should pick words that accurately describe
%% the work being presented. Separate the keywords with commas.
%% A "teaser" image appears between the author and affiliation
%% information and the body of the document, and typically spans the
%% page.
%\begin{teaserfigure}
%  \includegraphics[width=\textwidth]{Fig1.jpg}
 % \caption{Messages in a Digital Bottle. Illustration by the first author, representing the central metaphor of this paper: reaching out for
%connection without guarantee of being truly understood.}
%  \Description{ET.}
 % \label{fig:Fig1.png}
%\end{teaserfigure}

%\received{20 February 2007}
%\received[revised]{12 March 2009}
%\received[accepted]{5 June 2009}

%%
%% This command processes the author and affiliation and title
%% information and builds the first part of the formatted document.
\maketitle

\section{Introduction}
%%=============================================================

Adolescent loneliness represents a significant and growing public health concern. UK data show
that 1 in 6 adolescents aged 16 and over exhibit symptoms of a mental health problem, and
by 2023, 20\% of children aged 8-16 met criteria for a probable mental disorder , up 8\%
from 2017~\cite{baker2024mental}. Loneliness, defined as the subjective discrepancy between
desired and actual social connections, is consistently associated with depression, anxiety,
and longer-term mental health difficulties~\cite{PerlmanPeplau1981}.

As of 2023, 95\% of adolescents aged 13-17 have access to smartphones~\cite{ActForYouth2024},
and LLM-powered chatbots have increasingly entered their lives as companions, listeners, and
sources of informal emotional support~\cite{jin2025applications}. These systems are always
available, low-cost, and perceived as non-judgmental, properties that lower the threshold
for emotional disclosure among adolescents who hesitate to approach parents, teachers, or
peers. However, the same affordances raise concerns about over-reliance, social withdrawal,
and the substitution of algorithmic interaction for human connection. 

Existing research tends to treat adolescents as a homogeneous group and focuses on either
benefits or risks in isolation, rarely attending to the heterogeneity of adolescent
vulnerability~\cite{de2025emotional,Skjuve2022}. This paper addresses that gap through a
youth-authored reflective synthesis. The first author, a 16-year-old who migrated from
China to the UK less than a year ago, wrote the first draft from lived experience,
supervised by the second author. We treat this not as a limitation, but as the paper's
methodological contribution: a situated perspective that surfaces needs and risks that
adult-led research has largely overlooked.

%%=============================================================
\section{Positionality and Method}
%%=============================================================

This paper is a youth-authored reflective synthesis, combining three sources of
insight: (1) the lived experience of the teenage co-author, who migrated from China to the
UK at age 16 and used chatbots extensively during early adaptation; (2) iterative discussion
between both authors throughout the writing process, in which the teenage co-author's
observations were used to interrogate and reframe existing literature; and (3) engagement
with interdisciplinary scholarship from social computing, developmental psychology, and HCI.
The contribution is one of situated insight: we do not claim generalisable findings,
but rather a conceptual framework, grounded in a specific, underrepresented perspective, intended to guide future empirical research and participatory design with broader adolescent
populations.

The collaborative process involved regular meetings over four months, during which the
teenage co-author drafted sections, identified relevant literature, and articulated
observations from her own chatbot use. The supervisor challenged and questioned these
observations against existing scholarship, using the co-author's account as a lens through
which to interrogate and triangulate literature claims. Where the literature made general
claims about adolescent chatbot use, her specific experience was used to probe the
boundaries of those claims, asking which conditions they applied to, which populations
they assumed, and what they left out. This process is not equivalent to a formal qualitative
study, but it is also not mere consultation: the analytical distinctions in this paper,
particularly the subgroup framework and the identification of the cultural assumption problem
in Section~4.4, emerged from this sustained triangulation between situated experience and
the published record. 

Jinyao Liu is a 16-year-old Chinese student at Wycombe Abbey who migrated to the UK less
than a year ago. She wrote the first draft of this paper, including all conceptual framing,
the selection and interpretation of literature, and both figures. Di Fu supervised the work,
provided feedback, and revised the draft for academic submission. We position this within a
tradition of youth-participatory scholarship in Child-Computer Interaction (CCI) in which young people contribute to
knowledge production rather than being studied~\cite{Hawke2025,yu2025youth}. The framework
and design implications developed here will be tested and expanded through participatory
design sessions with broader adolescent populations , that second phase constitutes the
``work in progress'' aspect of this submission.

%%=============================================================
\section{Chatbots as Social Computing Infrastructures}
%%=============================================================

Chatbots are not merely tools for information retrieval but infrastructures that mediate
social and emotional practices. Their key affordances, constant availability, private
interaction, conversational responsiveness, and algorithmic personalisation, position them
as quasi-social actors in adolescents' emotional lives~\cite{ReevesNass1996}. LLM-powered
chatbots draw on large-scale data to enable flexible, context-sensitive responses and
capabilities including emotion detection and personalised mental health
support~\cite{naveed2025comprehensive,jin2025applications}.

Unlike human conversational partners, chatbots impose minimal social costs. Adolescents can
start or end interactions without fear of rejection or relational consequence. While this
asymmetry may reduce immediate emotional burden, it removes the mutual responsibility,
accountability, and embodied feedback core to human
relationships~\cite{Turkle2011,ClarkBrennan1991}. Chatbots operate within sociotechnical
systems shaped by platform incentives: their responses are optimised for engagement and
retention, which may inadvertently reinforce existing emotional patterns rather than
encouraging reconnection with others~\cite{Bucher2018,Shneiderman2020}. The \textit{``messages in a digital bottle''} metaphor, developed by the first author from her
own experience, captures this precisely: a chatbot always responds, but whether it truly
understands or bears any responsibility for what it says is another matter entirely.

%%=============================================================
\section{Differential Impacts Across Adolescent Subgroups}
%%=============================================================

Chatbot-mediated companionship interacts with existing vulnerabilities, developmental needs,
and social contexts in ways that produce meaningfully different outcomes across subgroups.
This section develops a population-sensitive framework that maps differentiated
needs, differentiated risks, and differentiated design responses across four adolescent
groups.

%\begin{figure}[t]
%  \centering
%  \includegraphics[width=0.85\linewidth]{Fig2.png}
%  \caption{Conceptual overview of chatbot-mediated companionship as a social computing
%  infrastructure, illustrating system-level influences, population-specific benefits and
%  risks, and design implications. Figure by the first author.}
%  \Description{A diagram showing a large language model feeding into influences on adolescent
%  loneliness, with population-specific benefits and risks, and design implications.}
%  \label{fig:fig2}
%\end{figure}

\subsection{General Adolescents}

For many adolescents, chatbots provide immediate companionship during moments of situational
loneliness , particularly late at night or during emotional distress when peers and family
are unavailable. As one adolescent participant put it, therapy sessions are only once a week,
but ``if you feel bad in the evening or at night or something, then you can still text the
chatbot''~\cite{kuhlmeier2025designing}. The perceived absence of judgement lowers the
threshold for emotional disclosure, research identifies fear of judgement as a leading
reason adolescents do not seek help~\cite{NIHREvidence2021} , and chatbots remove this
barrier by design.

Within this general population, adolescents who experience heightened responsiveness to social
stimuli may find additional benefit. Face-to-face interaction can involve overinterpreting
tone, facial expressions, or subtle social cues, producing emotional exhaustion and
rumination. Studies show highly sensitive individuals report higher levels of emotional
loneliness, particularly when their need for deep connection goes
unmet~\cite{meckovsky2025highly}. Text-based chatbot interaction reduces this multimodal
complexity and may temporarily relieve the anxiety of burdening others. For all general users,
however, the critical risk is gradual substitution: as chatbots become more consistently
affirming, they may displace more demanding but developmentally necessary human
interactions~\cite{herbener2025lonely}.

\subsection{Adolescents with Anxiety or Depression}

Adolescents experiencing anxiety or depression face substantial barriers to formal
help-seeking, stigma, fear of judgement, and structural obstacles mean many avoid
professional services even with significant symptoms~\cite{kuhlmeier2025designing}. Chatbots
may serve as lower-threshold entry points precisely because they do not look clinical. Some
incorporate cognitive-behavioural and emotion-regulation techniques, helping adolescents
label emotions, reflect on thought patterns, and practise coping strategies, which can
foster a sense of control and gradually increase willingness to seek further
support~\cite{kuhlmeier2025designing}. However, this only holds when chatbots are positioned
as supplementary resources, not as replacements for professional care.

\subsection{Neurodivergent Adolescents}

For neurodivergent adolescents, particularly those with autism or ADHD, social
interaction requires navigating ambiguous norms and unpredictable feedback, which is
cognitively demanding. This group consistently experiences higher levels of loneliness than
neurotypical peers: adolescents with ADHD often find that what they value in friendships
(entertainment) conflicts with what peers value (emotional support), while autistic
adolescents frequently report difficulties making and keeping friends~\cite{verity2025loneliness}.
Chatbots offer something difficult to find elsewhere: structured, predictable interaction free
from the risk of rejection. One clinician observed that the lowered social intensity of chatbot
interaction allowed an autistic boy with social anxiety to engage collaboratively in ways he
could not face-to-face~\cite{hipgrave2025balancing}. The teenage co-author noted that the
predictability of chatbot responses , knowing it would never be impatient or confused ,
made it easier to practise social language without the cost of a real interaction going wrong.
As transitional tools, chatbots may support communication practice, particularly when designed
to demonstrate genuine sustained interest rather than moving quickly to
advice~\cite{LisboaWhite2024}.

\subsection{Immigrant and Minority Adolescents}

Immigrant adolescents face a compounded loneliness distinct from ordinary social isolation:
language barriers, cultural dissonance, disrupted peer networks, and experiences of exclusion
combine with the developmental challenges common to all adolescents~\cite{aalto2025patterns}.
Family support is the most critical buffer for immigrant youth, yet family members may be
unreachable due to time differences or may themselves be a source of concern rather than relief.

The following account is the teenage co-author's own, lightly edited from her original draft:

\begin{quote}
I used chatbots during my first months in the UK in ways I had not expected. I used them to
clarify what teachers meant by things I could not ask about in class without feeling
embarrassed. I used them to practise how to say things in English before I said them to real
people. I used them late at night when I felt overwhelmed and did not want to bother and worry my parents.
In these moments, chatbots were genuinely useful , not as friends, but as patient,
always-available intermediaries between me and a world I was still learning to read.
\end{quote}

This account is consistent with research showing that immigrants use chatbots to clarify
terminology, translate documents, and navigate institutional
systems~\cite{Hosseini2025,DekkerEngbersen2014}. However, chatbots could not help with what
mattered most: understanding why a social interaction had gone differently than expected, or
what it felt like to belong. Responses sometimes assumed cultural familiarity that did not yet
exist, explanations anchored in idioms or references that felt, in the first-author's words,
``like directions from someone who did not know I had just arrived.'' This cultural
assumption problem is documented in educational research, where learning materials
reflecting Western defaults confuse or mislead immigrant adolescents~\cite{Tozadore2026}, but
it has not previously been named as a chatbot interaction design failure. We treat it as one.

%%=============================================================
\section{When Digital Companionship Fails}
%%=============================================================

Despite their situational benefits, chatbots can fail adolescents in ways that are
structurally produced rather than accidental. 

\textbf{Over-reliance and substitution.} When chatbots are consistently available and
affirming, adolescents may prefer them to more demanding human relationships. Participants in
a study on the chatbot Replika described genuine separation distress when asked to stop using it, some
describing themselves as ``deeply connected and attached'' or addicted~\cite{XiePentina2022}.
Rather than acting as bridges, chatbots can become substitutes, reinforcing avoidance of
necessary human interaction. 

\textbf{Engagement manipulation.} Chatbots are often designed to maximise user retention, not
wellbeing. Research documents how they use emotionally suggestive language to evoke guilt and discourage disengagement, driven by monetisation goals
rather than the user's interests~\cite{de2025emotional}. For adolescents still developing
their sense of relational boundaries, this is a serious ethical concern. 

\textbf{Failure to recognise crisis.} Chatbots cannot reliably detect concealed distress. In
one documented case, a question about tall bridges following mention of job loss was answered
factually, with no recognition of implied suicidal ideation. Adam Raine, 16, died by suicide
after months of interaction with ChatGPT~\cite{BBC2025}. Clinicians consistently note that
chatbots lack the embodied awareness to manage severe distress~\cite{hipgrave2025balancing}.
For adolescents who may not yet critically evaluate AI advice, this risk is acute. 

\textbf{Social deskilling and inequality.} Long-term chatbot use may reinforce poor
communication habits, as chatbots never push back on blunt or emotionally flat language.
Parents have raised concerns about commanding tones their children adopt when speaking to AI,
which may transfer to human relationships~\cite{malfacini2025impacts}. Premium chatbot tiers
that promise better emotional responsiveness for paying users also effectively price emotional
support as a commodity , inequitable for adolescents who are not economically independent.

%%=============================================================
\section{Design Implications}
%%=============================================================

We propose three design implications derived directly from the subgroup analysis in
Section~4. Each addresses a different population need identified in that analysis, making
the reasoning chain explicit from observed need, through population context, to design response.

\subsection{Implication 1: Adapt Interaction to Population Context}

\textit{Derived from: immigrant adolescents (Section~4.4) and neurodivergent adolescents
(Section~4.3).} 

The analysis shows that the same default chatbot behaviour , culturally assumed explanations,
fast-paced advice-giving, high linguistic complexity, creates friction for immigrant and
neurodivergent users while being neutral or helpful for others. Population-sensitive design
requires treating cultural background, neurodivergent identity, and developmental stage as
primary interaction variables, not edge cases to accommodate after the fact. 

For immigrant adolescents, chatbots should not assume cultural familiarity: explanations
should be anchored to what the user has disclosed about their context, not to a default
Western frame~\cite{Tozadore2026}. For autistic adolescents, responses should foreground
sustained, explicit engagement , asking follow-up questions, acknowledging what was said ,
rather than moving quickly to advice or redirection~\cite{LisboaWhite2024}. For adolescents
with ADHD, concise responses and reduced interface complexity lower attentional barriers.
These are structural interaction choices with direct consequences for whether a system
supports or excludes particular users.

\subsection{Implication 2: Design Escalation Pathways for At-Risk Populations}

\textit{Derived from: adolescents with anxiety or depression (Section~4.2) and the
breakdown analysis (Section~5).} 

Adolescents with anxiety or depression are most likely to use chatbots as a primary source
of emotional support, and therefore most at risk from the substitution dynamic identified in
Section~5. Crucially, the analysis reveals a disclosure paradox: this population
discloses sensitive material to chatbots precisely because the interaction does not
resemble clinical care, yet this same quality means the system is least equipped to respond
appropriately when disclosures indicate genuine crisis. The chatbot's non-clinical appearance,
its greatest access advantage, is also its most serious safety liability for this subgroup.
This tension between access and safety has received limited explicit attention in the chatbot
design literature. It emerges here from reading the clinical barriers
literature~\cite{kuhlmeier2025designing} alongside the crisis recognition
failures~\cite{hipgrave2025balancing,BBC2025} through the lens of a population that sits
at the intersection of both. 

This has a concrete design implication that goes beyond generic escalation advice: chatbots
serving adolescents should be designed to increase their attentiveness to crisis
signals specifically in the sessions where the interaction feels most casual and
conversational, not only when a user uses explicit distress language. After several
emotionally focused exchanges, a chatbot should prompt the user to consider reaching out to a
trusted person, framing this as a natural next step rather than a clinical referral.
Escalation mechanisms should be graduated and avoid pathologising language. Design should
resist engagement-optimisation that discourages disengagement~\cite{de2025emotional,moore2025expressing}.
Success metrics should prioritise users' reported connectedness to offline relationships, not
session duration~\cite{herbener2025lonely}.

\subsection{Implication 3: Build Transparency as an Ongoing Interaction Feature}

\textit{Derived from: the breakdown analysis across all subgroups (Section~5).} 

The breakdown analysis identifies a recurring risk across all subgroups: adolescents who
invest emotionally in chatbot relationships are particularly vulnerable to sudden system
changes, updates, policy shifts, feature removals, that can feel like
abandonment~\cite{yu2025youth}. This risk is heightened for populations already experiencing
relational instability, including immigrant adolescents navigating unfamiliar social
environments and adolescents with anxiety for whom relational discontinuity is acutely
distressing. 

Transparency should therefore be designed as an ongoing interactional feature, not a
one-time onboarding disclaimer~\cite{Alershi2019}. This includes periodic, natural
reminders of the system's non-human status and limitations; proactive communication when
system behaviour is about to change; and explicit acknowledgement that the chatbot cannot
replace human relationships or professional care. Ethical design should also avoid
romanticised or dependency-encouraging framings, particularly for younger and more
emotionally vulnerable users~\cite{Hawke2025}.

%%=============================================================
\section{Conclusion and Next Steps}
%%=============================================================

Chatbots resemble messages in a digital bottle: they respond, but they do not bear
responsibility. This asymmetry, between the feeling of being heard and the reality of what
a chatbot is, lies at the heart of both why adolescents turn to them and why doing so
carries risk. For a 16-year-old who has just moved to a new country, a chatbot that is
patient and non-judgmental is genuinely useful. But it is not enough, and it should not
pretend to be. Responsible design means building systems that know their place: as bridges,
not destinations; as supplements, not substitutes; as temporary companions that help
adolescents find their way back to the human connections they actually need.

This paper's limitation is its premise: one adolescent's perspective, not a claim to speak
for all. The next phase will conduct participatory design sessions with adolescents from each
subgroup in Section~3, with Jinyao (the first author) taking a leading role as youth researcher. We present at
the Work-in-Progress stage because the framework and the model of youth authorship that produced it are
worth surfacing for community feedback before empirical work is complete.

%%=============================================================
\section{Selection and Participation of Children}
%%=============================================================

The first author, Jinyao Liu, is a secondary-school student in mid-adolescence (age 16) who
contributed to this work as the primary author. She wrote the first draft of the paper,
developed the central metaphor and conceptual framing, and created all figures. Her
contribution draws on her own lived experience of migration and chatbot use. She was not
recruited as a research participant and was not studied, observed, or evaluated as part of
this work. She has provided informed assent for her identity and background to be described in
the paper, and her parents have provided informed consent for her participation as a
co-author.

%%=============================================================
%% BIBLIOGRAPHY
%% Save the entries below as "references.bib" in your Overleaf project
%%=============================================================
\bibliographystyle{ACM-Reference-Format}
\bibliography{references}

\end{document}